\documentclass[aps,prl,twocolumn,preprintnumbers,nofootinbib,superscriptaddress,altaffilletter]{revtex4}
\usepackage{graphicx}
\usepackage{amssymb}
\usepackage{amsmath,mathrsfs,verbatim}
\usepackage{times}
\usepackage{latexsym}
\def\beq{\begin{equation}}
\def\eeq{\end{equation}}
\def\bey{\begin{eqnarray}}
\def\eey{\end{eqnarray}}

\def\lsim{\mathrel{\raise.3ex\hbox{$<$\kern-.75em\lower1ex\hbox{$\sim$}}}}
\def\gsim{\mathrel{\raise.3ex\hbox{$>$\kern-.75em\lower1ex\hbox{$\sim$}}}}

\newcommand{\be}{\begin{equation}}
\newcommand{\ee}{\end{equation}}

\begin{document}

\preprint{MCTP-12-06}

\title{Confronting Top $A_{FB}$ with Parity Violation Constraints}

\author{Moira I. Gresham}
\affiliation{Whitman College, Walla Walla, WA 99362}
\author{Ian-Woo Kim}
\affiliation{Michigan Center for Theoretical Physics, University of Michigan, Ann Arbor, MI 48109}
\author{Sean Tulin}
\affiliation{Michigan Center for Theoretical Physics, University of Michigan, Ann Arbor, MI 48109}
\author{Kathryn M. Zurek}
\affiliation{Michigan Center for Theoretical Physics, University of Michigan, Ann Arbor, MI 48109}

\date{\today}

\begin{abstract}

We consider the implications of low-energy precision tests of parity violation on $t$-channel mediator models explaining the top $A_{FB}$ excess measured by CDF and D0.  Flavor-violating $u$-$t$ or $d$-$t$ couplings of new scalar or vector mediators generate at one-loop an anomalous contribution to the nuclear weak charge.  As a result, atomic parity violation constraints disfavor at $\gtrsim 3 \sigma$ $t$-channel models that give rise to a greater than 20\% $A_{FB}$ at the parton level for $M_{t\bar t}> 450$ GeV while not producing too large a $t\bar t$ cross-section.   Even stronger constraints are expected through future measurements of the proton weak charge by the Q-Weak experiment.

\end{abstract}
\maketitle

\paragraph{Introduction:}  As the heaviest particle in the Standard Model (SM), the top quark provides a special window into new physics at the electroweak symmetry breaking scale.  In fact, the most persistent anomaly to come from the Tevatron arises in the top system. Both the CDF and D0 collaborations have reported an excess in measurements of the $t \bar t$ forward-backward asymmetry $A_{FB}$, favoring production of $t$ in the incoming proton direction, and $\bar t$ in the incoming antiproton direction.  CDF observed $A_{FB} = 0.475 \pm 0.114$ for $t \bar t$ invariant mass $M_{t\bar{t}} > 450$ GeV~\cite{Aaltonen:2011kc} at the parton level ($A_{FB} = 0.266 \pm 0.062$ at the signal level), a 3.4$\sigma$ deviation from the SM next-to-leading order (NLO) prediction of $0.088 \pm 0.013$ ($0.043 \pm 0.009$ at the signal level).  D0 has confirmed the $A_{FB}$ excess, though without the dramatic rise at the high $M_{t\bar{t}}$ \cite{Abazov:2011rq}.  At the signal level, within errors, the two experiments agree with each other.

Most new physics models that may account for this excess fall into two classes: $s$-channel and $t$-channel. The $s$-channel models involve a new colored resonance with axial couplings ({\it e.g.}, axigluons)~\cite{Frampton:2009rk,Sehgal:1987wi,Ferrario:2009bz}, although the simplest  such models have become disfavored due to the absence of $t\bar{t}$ resonances at high invariant mass at the LHC~\cite{LHCnotes}.   The $t$-channel models feature a scalar or vector mediator, denoted $M$, with a flavor-violating coupling $\lambda$ between $u$ or $d$ and $t$ (or $\bar t$), and can generate a large forward-backward asymmetry through a Rutherford enhancement~\cite{Jung:2009jz,Dorsner:2009mq}.  Heavy mediators $(m_M >m_t)$ have become disfavored by the invariant mass distribution and number of additional jets in $t \bar t$ events at the LHC~\cite{LHCnotes}, due to a large $t \bar t + \rm{jet}$ cross section from on-shell $M$ production~\cite{Gresham:2011fx,Jung:2011zv}.  Light mediators ($m_M < m_t$) therefore are the most promising for evading collider constraints; on-shell $M$ production does not contribute to $t \bar t$ since $M$ cannot decay to $t+\rm{jet}$.

\begin{figure}[b]
\begin{center}
\includegraphics[scale=.8]{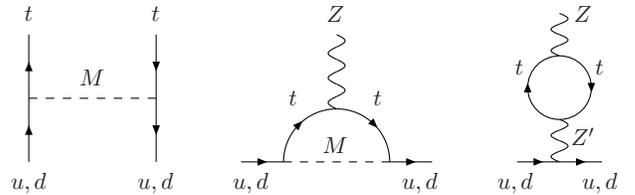}
\caption{\it $A_{FB}$ from $t$-channel exchange of $M$ (left).  Anomalous coupling of $Z$ to $u,d$ at one-loop is generated by $M$ (center) and by flavor-conserving $Z^\prime$ associated with certain vector $M$ models.
 \label{feyn}}
\end{center}
\end{figure}

In this Letter, we show that low-energy precision tests of parity-violating (PV) observables disfavor $t$-channel models for $A_{FB}$.  As shown in Fig.~\ref{feyn}, all $t$-channel models generically lead to an anomalous coupling of the $Z$ boson to $u$ or $d$ quarks at one-loop, which is of order $(\lambda^2/(4\pi)^2) \, (m_t^2/m_M^2) \sim 10^{-2}$,  for $\lambda \sim 1$ and $m_M \sim m_t$ in order to explain $A_{FB}$.  Atomic parity violation (APV) measurements in cesium~\cite{Wood:1997zq} provide the strongest constraints, at the level of $10^{-3}$, and the upcoming proton weak charge measurement by the Q-Weak experiment~\cite{VanOers:2007zz} is expected to give even stronger limits.  We emphasize that PV measurements are particularly sensitive to $t$-channel models with light mediators, therefore providing a complementary test of models for $A_{FB}$ that are most easily hidden in collider searches.  We consider here simple scalar and vector $t$-channel models, which have thus far evaded collider bounds, and find that they are strongly excluded by PV constraints.


\paragraph{Parity violation constraints:} PV electron-quark interactions can be parametrized below the weak scale by an effective four-fermion interaction
\be
\mathscr{L}_{eq}^{PV} = \frac{G_F}{\sqrt{2}} \sum_{q = u,d} \big( C_{1q} \bar{e} \gamma^\mu \gamma_5 e \, \bar{q} \gamma_\mu q + C_{2q} \bar{e} \gamma^\mu  e \, \bar{q} \gamma_\mu \gamma_5 q \big) \,  . \label{PVeq}
\ee
In the SM, the coefficients $C_{1q}$ and $C_{2q}$ arise at leading order via $Z$ exchange: {\it e.g.}, $C_{1u} = - \frac{1}{2} + \frac{4}{3} s_W^2$ and $C_{1d} = \frac{1}{2} - \frac{2}{3} s_W^2$, where $s_W \equiv \sin \theta_W$ describes the weak mixing angle.  Beyond leading order, precision SM computations~\cite{Marciano:1982mm,Erler:2004in} allow for stringent constraints on new physics contributing to Eq.~\eqref{PVeq}, denoted $C_{1q}^{\rm NP}$ and $C_{2q}^{\rm NP}$.

APV experiments provide the most precise measurements of $C_{1q}$.  Interference between $\gamma$ and $Z$ amplitudes give rise to PV atomic transitions sensitive to the nuclear weak charge
\be
Q_W(Z,N) \equiv -2\big[ (2Z+N)C_{1u} + (2N+Z)C_{1d}\big] \; .
\ee
The strongest constraint is from cesium ($^{133}$Cs)~\cite{Wood:1997zq}, for which the measured value $Q_W(\textrm{Cs}) = -73.20(35)$ agrees with the SM prediction $Q_W^{\rm SM}(\textrm{Cs}) = -73.15(2)$~\cite{Porsev:2009pr,Nakamura:2010zzi},
probing $C_{1q}^{\rm NP}$ at the ${\rm few} \times 10^{-3}$ level. (Uncertainty in the last digits is given in parantheses.)

Another constraint on Eq.~\eqref{PVeq} is provided by the proton weak charge $Q_W(p)$ measured in parity-violating $e$-$p$ elastic scattering (see \cite{Roche:2011zz} and references therein).  Ref.~\cite{Young:2007zs} obtained $Q_W(p) = 0.054(17)$, in $1\sigma$ agreement with the SM value $Q_W^{\rm SM}(p) = 0.0713(8)$.
The new physics reach in $Q_W(p)$~\cite{Erler:2003yk} will be dramatically improved by the Q-Weak experiment~\cite{VanOers:2007zz}, which aims to measure $Q_W(p)$ to $4 \%$, corresponding to a $10^{-3}$ sensitivity to $C_{1q}^{\rm NP}$.

We consider new physics models, described below, that generate anomalous couplings of the $Z$ to light quarks $q=u,d$, given by
\be
\mathscr{L}_{\rm eff} = - \frac{g_2}{c_W}  Z^\mu \left( a_{R}^{\rm NP}(q) \, \bar{q}_{R} \gamma_\mu q_{R}+a_{L}^{\rm NP}(q)  \, \bar{q}_{L} \gamma_\mu q_{L} \right) \label{LNP}
\ee
where $a^{\rm NP}_{L,R}(q)$ parametrizes the new physics contribution.  Constraints on these couplings from the hadronic $Z$ width were considered previously in connection with $A_{FB}$~\cite{Grinstein:2011dz}, but are weaker than those from APV.  In terms of Eq.~\eqref{PVeq}, we have $C_{1q}^{\rm NP} = a^{\rm NP}_{L}(q) + a^{\rm NP}_{R}(q)$ and $C_{2q}^{\rm NP} = Q_W(e) [a^{\rm NP}_{R}(q) - a^{\rm NP}_{L}(q)]$.  We do not consider $C_{2q}^{\rm NP}$ since it is suppressed by the electron weak charge $Q_W(e) \approx (-1+4 s_W^2) \approx - 0.04$.

Additional constraints on Eq.~\eqref{LNP} arise from neutrino deep inelastic scattering ($\nu$DIS) experiments~\cite{Paschos:1972kj}.  The low-energy $\nu$-$q$ interaction can be parametrized as
\begin{align}
\mathscr{L}_{\nu q}^{PV} &=  - \frac{G_F}{\sqrt{2}} \sum_{q = u,d} \bar{\nu} \gamma^\mu (1- \gamma_5) \nu \notag \\
& \;\times   \big( \epsilon_L(q) \, \bar{q} \gamma_\mu (1- \gamma_5) q + \epsilon_R(q) \, \bar{q} \gamma_\mu (1+ \gamma_5) q \big) 
\end{align}
where $\epsilon_R(u) = \epsilon_L(u) - \frac{1}{2} = - \frac{2}{3} s_W^2$ and $\epsilon_R(d) = \epsilon_L(d) + \frac{1}{2} = \frac{1}{3} s_W^2$ at leading order in the SM.  The quantites $g_L^2 \equiv \sum_q \epsilon_L^2(q) = 0.3025(14)$ and $g_R^2 \equiv \sum_q \epsilon_R^2(q) = 0.0309(10)$ measured in neutral-to-charged-current ratios of $\nu$ and $\bar{\nu}$ cross sections on isoscalar nuclear targets agree with SM predictions $(g_L^2)_{\rm SM} = 0.30499(17)$ and $(g_R^2)_{\rm SM}=0.03001(2)$~\cite{Nakamura:2010zzi},
constraining any NP contribution $\epsilon_{L,R}^{\rm NP}(q) =- a_{L,R}^{\rm NP}(q)$.
Since $a_{L,R}^{\rm NP}$ enters predominantly via interference with the SM couplings $\epsilon_{L,R}$, $\nu$DIS gives weaker constraints on right-handed couplings.


\paragraph{New physics models for top $A_{FB}$:} We consider a set of simple models, given in Table~\ref{models}, to generate $A_{FB}$ through $t$-channel exchange of a scalar or vector mediator.  We focus on mediators coupling $t$ to $u_R$ only, thereby generating $a_{R}^{\rm NP}(u)$. 
Other models with couplings to $(u,d)_L$ or $d_R$ generate $a_{L}^{\rm NP}(u,d)$ or $a_{R}^{\rm NP}(d)$, respectively; the former case requires an extended flavor-symmetric new physics sector~\cite{Grinstein:2011dz} to avoid constraints from $K^0$-$\bar{K}^0$ or $D^0$-$\bar{D}^0$ mixing~\cite{Blum:2011fa}, and the latter suffers from smaller parton luminosity, requiring larger couplings.  In any case, APV is equally sensitive to all $a_{L,R}^{\rm NP}(u,d)$ since Cs is approximately isoscalar.

\begin{table}[b!]
\begin{center}
\begin{tabular}{|c|c|}
\hline
New mediator field& Interaction Lagrangian $\mathscr{L}_{\textrm{int}}$ \\
\hline
scalar $\phi \sim (1,2,1/2)$ & $\lambda\, ( \bar u_R V_{ib} u^i_L \phi^0 - \bar u_R b_L \phi^+)+ \textrm{h.c.}$ \\
scalar $\omega \sim (3,1,-4/3)$ & $\lambda \,\epsilon_{\alpha\beta\gamma} \bar{t}_{R\alpha}^c u_{R\beta} \, \omega_\gamma+ \textrm{h.c.}$ \\
vector $V^\prime \sim (1,1,0)$ & ${\lambda} \, \bar{t}_R \gamma^\mu u_R V^\prime_\mu+ \textrm{h.c.}$ \\
\hline
\end{tabular}
\caption{\it New states and interactions introduced to explain $A_{FB}$ via $t$-channel exchange, with real coupling constant $\lambda$.  $SU(3) \times SU(2)_L \times U(1)_Y$ quantum numbers are given in parantheses.  \label{models}}
\end{center}
\end{table}


In order to calculate $A_{FB}$, $\sigma(t\bar{t})_{\ell j}$, and $\sigma(t\bar{t})_{\ell \ell}$ at leading order (LO) for $t \bar{t}$+0,1 jet samples within new physics models, events were generated using MadGraph/MadEvent 5 v1.3.32 \cite{Alwall:2011uj} and Pythia v6.420. MLM Matching, a fixed RG scale of 200 GeV, $m_t $=172 GeV, and CTEQ6L1 parton distribution functions were used.  Model files were generated  using FeynRules v1.6.0.  $10^5$ events were generated for an array of mass and coupling values for each model. Contours were generated by interpolating between model points that saturated the given bounds.


For scalar mediators, we consider color triplet $(\omega)$ diquarks \cite{Shu:2009xf,Arhrib:2009hu}
and a color singlet, weak doublet $\phi \!=\! (\phi^+, \phi^0)$ \cite{Nelson:2011us,Babu:2011yw,Grinstein:2011dz,Blum:2011fa}.  The latter model, for $m_{\phi^0} \lesssim 130$ GeV, has been argued to provide the best fit among scalar mediators for $A_{FB}$ and other constraints~\cite{Blum:2011fa}, while potentially accounting for flavor anomalies~\cite{Hochberg:2011ru,Ng:2011jv}.
For these mediators $M=\phi^0,\, \omega$, the new physics coefficient is 
\be
a^{\rm NP}_R(u) = \frac{\lambda^2 c_M}{32 \pi^2} \,\frac{m_t^2 }{m_M^2} F( m_t^2 / m_M^2 ) 
\ee
where $F(x) \equiv (x - 1 - \log x)/(1-x)^2$, and $c_\phi = |V_{tb}|^2 \approx 1$, $c_\omega = 2$.  (The $\phi$ result is independent of the $\phi^+$ mass for $m_{\phi^+} \gg m_b$.)

\begin{figure}[t]
\begin{center}
\includegraphics[scale=.6]{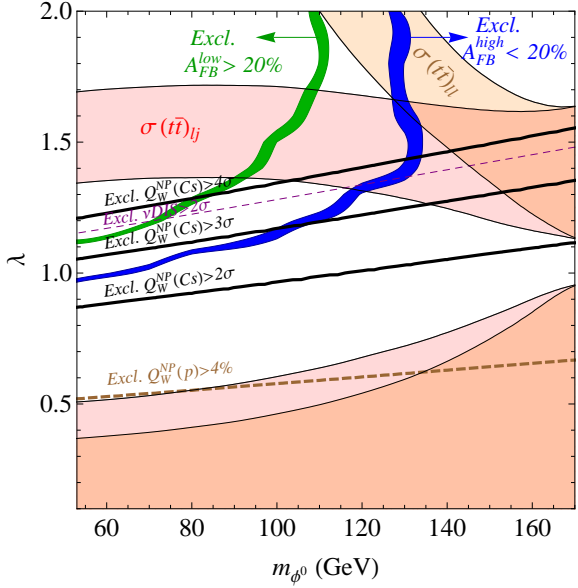} 
\caption{\it Exclusion plot for weak doublet ($\phi$) model.  Pink and tan shaded regions are consistent with $\sigma(t \bar t)_{\ell j}$ and $\sigma(t \bar t)_{\ell \ell}$, respectively.  Mass-dependent-$A_{FB}$-favored region is within the blue and green curves, marking $A_{FB}^{\rm high} > 20\%$ and $A_{FB}^{\rm low} < 20\%$, respectively. Constraints from $Q_W({\rm Cs})$, $\nu$DIS, and future $Q_W(p)$ measurements shown by black solid, purple dashed, and brown dashed lines, respectively. \label{contours}}
\end{center}
\end{figure}

Our results for the weak doublet model are shown in Fig.~\ref{contours}.  The blue and green lines show the preferred region for $A_{FB}$, given at the parton level and including only new physics contributions, in the high ($M_{t \bar t} > 450$ GeV)  and low ($M_{t \bar t} < 450$ GeV) invariant mass bins, respectively.   We impose $A_{FB}^{\rm high} > 20\%$ and $A_{FB}^{\rm low} < 20\%$.  The line thickness corresponds to statistical uncertainty in our simulation. The total $t \bar t$ cross section $\sigma(t \bar t)$ has been measured at CDF in semileptonic ($\ell j$) and dileptonic ($\ell \ell$)  channels (where $\ell = e,\mu$), both in agreement with SM prediction \cite{Aaltonen:2010bs,Aaltonen:2010ic}.  We require $\sigma(t\bar t)$ agree with SM prediction at LO within $\pm 30\%$ in each channel, shown by the shaded regions; this large uncertainty reflects our ignorance of acceptance effects, NLO corrections, and uncertainties in the cross-section and top mass measurements.  
The $\phi^0$ modifies $\sigma(t \bar t)_{\ell j}$ and $\sigma(t \bar t)_{\ell \ell}$ through both $t \bar t$ production and $t$ decays, since $t \to \phi^0 u$ is allowed (with $\phi^0$ decaying hadronically via Cabibbo-suppressed coupling to $\bar{u}_R c_L$).  Interference between QCD and $\phi^0$-mediated $t \bar t$ production is destructive, requiring a large $\mathcal{O}(\lambda^4)$ new physics-squared contribution to compensate.  Moreover, $\sigma(t \bar t)_{\ell \ell}$ is further suppressed, compared to $\sigma(t \bar t)_{\ell j}$, by the reduced leptonic branching ratio, requiring larger values of $\lambda$ and leading to a tension between $\sigma(t \bar t)_{\ell \ell}$ and $\sigma(t \bar t)_{\ell j}$.

The constraints from low-energy PV observables, shown in Fig.~\ref{contours}, clearly exclude the weak doublet model as the origin of $A_{FB}$.  The $Q_W({\rm Cs})$ and $\nu$DIS measurements disfavor this model at $4\sigma$ (solid line) and $2 \sigma$ (dashed line), respectively.  The Q-Weak measurement of $Q_W(p)$ can provide even stronger constraints (thick dashed line).  PV constraints similarly disfavor the diquark models.  In Table~\ref{bench}, we list a couple of diquark benchmark points that provide reasonable agreement with $A_{FB}$ and $\sigma(t \bar t)$, but give a large disagreement with PV measurements.  

\begin{figure}[t]
\begin{center}
\includegraphics[scale=.60]{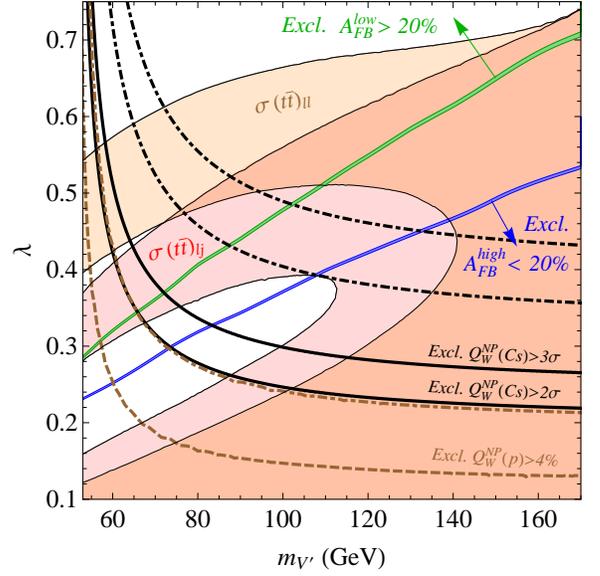}
\caption{\it Exclusion plot for horizontal $SU(2)_X$ $V^\prime$ model, as in Fig.~\ref{contours}.  Constraints from $Q_W({\rm Cs})$ and future $Q_W(p)$ measurements shown by solid black and brown dashed lines, respectively, from Eq.~\eqref{naive}. Dot-dashed lines show same constraints from Eq.~\eqref{full}.\label{contours2}}
\end{center}
\end{figure}

\begin{table}[b!]
\begin{center}
\begin{tabular}{|c |c c c c c c c |}
\hline
\, scalar \, & $m_M$ & $\lambda$ & $A_{FB}^{\rm high}$ & $\sigma(t \bar t)_{\ell j}$ & $a_R^{\rm NP}(u)$ & $Q_W^{\rm NP}({\rm Cs})$ & $Q_W^{\rm NP}(p)$\\
\hline
$\omega$ & $600$& $3.5$& $25\%$ & $7.0$ pb & $0.012$ & $-4.5$ & $-0.05$\\
& $800$ & $4.2$ & $26\%$ & $6.7$ pb & $0.012$ & $-4.5$ & $-0.05$ \\
\hline
$\phi^0$ & $130$ & $1.6$ &  $20\%$ & $7.4$ pb & $0.0048$ & $-1.8$ & $-0.02$\\
\hline
$V^\prime$ & $160$ & $0.55$ & $30\%$ & $5.1$ pb & $0.012$ & $-4.6$ & $-0.05$\\
\hline
\end{tabular}
\caption{\it Benchmark points: (i) color triplet diquark $\omega$; (ii) weak doublet $\phi=(\phi^+,\phi^0)$, similar to the ``best-fit'' point of Ref.~\cite{Blum:2011fa} (our $\lambda$ convention differs by factor 2); and (iii) horizontal $V^\prime$ model, similar to ``Model A'' point of Ref.~\cite{Jung:2011id}, with PV coefficients computed using Eq.~\eqref{naive}, $m_{Z^\prime} = 120$ GeV, $\Lambda=600$ GeV. LO $\sigma(t \bar t)_{\ell j}$ should be compared to $\sigma(t \bar t)_{\ell j}^{\rm SM} = 6.3$ pb at LO.
\label{bench}}
\end{center}
\end{table}

Next, we consider models with a vector mediator, denoted $V^\prime$, coupled to $t_R$-$u_R$.  We focus on the model of Ref.~\cite{Jung:2011zv}: the SM is extended with an $SU(2)_X$ horizontal symmetry acting on $(u,t)_R$, giving rise to a complex $V^\prime$ and a real, flavor-conserving $Z^\prime$, analogous to the SM $W$ and $Z$.  The fermion-gauge interactions are
\begin{align}
&\mathscr{L} = \frac{g_X}{\sqrt 2} V^\prime_\mu \big[  \bar{u}_R \gamma^\mu t_R + \varepsilon ( \bar u_R \gamma^\mu u_R - t_R \gamma^\mu t_R ) \big]  + \textrm{h.c.}\notag  \\
& + \frac{g_X}{2} Z_\mu^\prime \big[ \bar t_R \gamma^\mu t_R - \bar u_R \gamma^\mu u_R  + 2 \varepsilon( \bar{u}_R \gamma^\mu t_R +  \bar{t}_R \gamma^\mu u_R) \big]
\end{align}
where $g_X \equiv \sqrt{2} \lambda$ is the gauge coupling and $\varepsilon$ corresponds to a vacuum misalignment between different $SU(2)_X$-breaking Higgs fields. We assume $\varepsilon \ll 1$, to avoid same-sign top production, and neglect $\mathcal{O}(\varepsilon^2)$ terms. The prefered region for collider constraints is: (i) $m_{V^\prime} < m_t$, such that on-shell $V^\prime$ production does not contribute to the $t \bar t$ sample, since $V^\prime \to u \bar u$ can dominate over $V^\prime \to u \bar{t}^*$ for $\varepsilon \ne 0$; and (ii) $m_{Z^\prime} \lesssim 130$ GeV to avoid dijet bounds ($m_{Z^\prime} \gtrsim \textrm{TeV}$ is also viable, but requires $\mathcal{O}(100)$-dimensional $SU(2)_X$ Higgs representations)~\cite{Jung:2011zv,Jung:2011id}.  This model generates $a_{R}^{\rm NP}(u)$, but it is not possible to compute $a_{R}^{\rm NP}(u)$  in a model-independent way since the theory is nonrenormalizable unless we specify how $SU(2)_X$ is spontaneously broken.
Nevertheless, we can obtain a reasonable estimate for $a_{R}^{\rm NP}(u)$ by assuming these degrees of freedom enter at scale $\Lambda$, and treating $\Lambda$ as a cut-off.  We find
\begin{align}
a_{R}^{\rm NP}(u) = &- \frac{\lambda^2}{16\pi^2} \frac{m_t^2}{m_{V^\prime}^2} \Big( F\Big(\frac{m_t^2}{m_{V^\prime}^2} \Big) + \frac{1}{4} \log \Big(\frac{\Lambda^2}{m_t^2}\Big) \Big) \notag   \\
&+ \frac{N_C \lambda^2}{32 \pi^2} \frac{m_t^2}{m_{Z^\prime}^2} \log \Big(\frac{\Lambda^2}{m_t^2} \Big) \, , \label{naive}
\end{align}
with the two terms corresponding to vertex and $Z$-$Z^\prime$ mixing contributions, respectively.

It is also useful to consider a specific ultraviolet completion of the $SU(2)_X$ model in which $a_{R}^{\rm NP}(u)$ can be computed.  In order to break $SU(2)_X$, we introduce two (SM singlet) scalar fields: a complex doublet $S$ and a real triplet $\Sigma$, with vacuum expectation values (vevs) taken to be $\langle S \rangle = (0,v_S)$ and $\langle \Sigma \rangle = v_\Sigma(-2\varepsilon, 0,1)/\sqrt{2}$.  We also introduce a massive vector quark $t^\prime \sim (3,1,2/3)$, which is a singlet under $SU(2)_X$, with mass $m_{t^\prime} \gg m_t$ and Yukawa interactions
\be
\mathscr{L} = y_1 (\bar u_R, \bar t_R) t^\prime_L S - y_2 \bar{t}_R^\prime (t_L,b_L) \epsilon H + \textrm{h.c.}
\ee
with antisymmetric tensor $\epsilon$.  The SM Higgs field is $H \equiv (H^+, H^0)$, with vev $\langle H^0 \rangle = v$.  Integrating out the $t^\prime$ generates the top mass $m_t = y_1 y_2 v_S v/m_{t^\prime}$.  While $S$ is required to generate $m_t$, $\Sigma$ is required to break the degeneracy between $m_{V^\prime}^2 = g_X^2 (v_S^2 + v_\Sigma^2)/2$ and $m_{Z^\prime}^2 = g_X^2 v_S^2/2$ and to generate $\varepsilon$.  (We neglect other SM quark masses.)  Within this concrete realization, we have
\begin{align}
a_{R}^{\rm NP}(u) = &- \frac{\lambda^2}{16\pi^2} \frac{m_t^2}{m_{V^\prime}^2} F_1\Big(\frac{m_t^2}{m_{V^\prime}^2}, \frac{m_{t^\prime}^2}{m_{V^\prime}^2} \Big) \notag \\
& + \frac{N_C \lambda^2}{32\pi^2} \frac{m_t^2}{m_{Z^\prime}^2} F_2\Big(\frac{m_t^2}{m_{t^\prime}^2}\Big) \; , \label{full}
\end{align}
with loop functions from vertex and $Z$-$Z^\prime$ mixing contributions, respectively, given by
\begin{subequations} 
\begin{align}
F_1(x,y) \equiv &- \frac{1}{4} \Big( 2 + \frac{6-3x-3y}{(1-x)(1-y)} \\
& + \frac{(x^2-2x + 4)\log x}{(1-x)^2} + \frac{(2x^2-8x)\log x}{(1-x)(x-y)} \notag \\
&+ \frac{(y^2-2y + 4)\log y}{(1-y)^2} + \frac{(2y^2-8y)\log y}{(1-y)(y-x)} \Big) \notag \\
F_2(x) \equiv &\frac{2(x-1) -(1+x) \log x}{1-x} \; .
\end{align}
\end{subequations}
In the $m_{t^\prime} \gg m_{V^\prime}, m_t$ limit, Eq.~\eqref{full} reproduces the $\log \Lambda$ dependence of  Eq.~\eqref{naive}, with $\Lambda \equiv m_{t^\prime}$.

In Fig.~\ref{contours}, we show that PV observables provide strong constraints on the $SU(2)_X$ model.  The preferred region for $A_{FB}$ lies between the blue and green curves, while $\sigma(t \bar t)_{\ell \ell}$ and $\sigma(t \bar t)_{\ell j}$ measurements favor the overlap of the shaded regions.  The solid curves show exclusion limits from APV measurements, and the dashed curve indicates the potential reach of the Q-Weak measurement, computed using Eq.~\eqref{naive} for $m_{Z^\prime} = 120$ GeV and $\Lambda = 600$ GeV.  The dot-dashed lines show the corresponding PV constraints for the complete model, using Eq.~\eqref{full}, with $m_{t^\prime} = \Lambda$.  The constraints become stronger for smaller values of $m_{Z^\prime}$ or larger values of $\Lambda$.  The difference in the limits obtained with Eqs.~\eqref{naive} and \eqref{full} gives a qualitative view of the model dependence in $a_{R}^{\rm NP}(u)$, and the agreement becomes much better for larger $\Lambda$.  For light $Z^\prime$, $Z$-$Z^\prime$ mixing dominates; for intermediate mass, $130 \, \textrm{GeV}< m_{Z^\prime} \lesssim 1$ TeV, there can be a cancellation between mixing and vertex terms, but this region is disfavored by dijet searches; for $m_{Z^\prime} \gtrsim 1$ TeV, the vertex terms dominate and $a_{R}^{\rm NP}(u)$ is comparable in size to the Q-Weak sensitivity.

\paragraph{Conclusions:} We studied in detail the most promising $t$-channel models for top $A_{FB}$. We showed that these models, characterized by a new color singlet, weak doublet scalar and a new color singlet, weak singlet vector of low mass, are strongly disfavored by PV constraints. More generally, we showed that \emph{any} low-mass $t$-channel model for top $A_{FB}$ will confront very strong bounds from atomic parity violation measurements.


{\it Acknowledgements:}  We thank Y.~Hochberg, A.~Pierce, and H.~B.~Yu for helpful discussions.   ST was supported by DOE Grant \#DE-FG02-95ER40899 and KZ was supported by NSF CAREER award PHY1049896.

\end{document}